\documentclass[12pt]{article}

\usepackage{graphicx}

\usepackage{verbatim} 
\usepackage{graphicx}
\usepackage{srcltx}
\usepackage{amsmath}
\usepackage{amsthm}
\usepackage{amsmath}
\usepackage{amsfonts}
\usepackage{amssymb}
\usepackage{amscd}
\usepackage{enumerate}
\usepackage{dsfont}
\usepackage{cite}
\usepackage{setspace}

\usepackage{appendix}

\usepackage{rotating}

\usepackage[automark]{scrpage2}

\def\C{\mathbb{C}}
\def\H{\mathcal{H}}

\def\p{\psi}

\def\2{\frac{1}{2}}

\def\Tr{\mathrm{Tr}}

\def\be{\begin{equation}}
\def\ee{\end{equation}}
\def\bp{\begin{proof}}
\def\ep{\end{proof}}
\def\bc{\begin{cases}}
\def\ec{\end{cases}}

\def\bea{\begin{eqnarray}}
\def\eea{\end{eqnarray}}

\newcommand{\bra}[1]{\ensuremath{\left\langle #1\right|}}
\newcommand{\ket}[1]{\ensuremath{\left|#1\right\rangle}}
\newcommand{\braket}[2]{\ensuremath{\left\langle #1\vphantom{#2}\right.\left|\vphantom{#1}#2\right\rangle}}

\begin{document}


\title{Tensorial characterization and quantum estimation of weakly entangled qubits}

\author{ P. Aniello$^{1}$, J. Clemente-Gallardo$^2$, G. Marmo$^1$, G. F. Volkert$^3$\\\\
{\emph {\small $^1$ Dipartimento di Scienze Fisiche dell'Universit\`a di Napoli ``Federico II'' and }}\\ 
{\emph {\small  INFN -- Sezione di Napoli, Complesso Universitario di Monte Sant'Angelo, }}\\ 
{\emph {\small via Cintia, I-80126 Napoli, Italy}}\\
{\emph {\small E-mail: aniello@na.infn.it, marmo@na.infn.it}}\\\\
{\emph {\small $^2$ BIFI and Departamento de F\'isica Te\'orica,}}\\  
{\emph {\small Edificio I+D--Campus R\'io Ebro, 50018 Zaragoza, Spain}}\\
{\emph {\small E-mail: jesus.clementegallardo@bifi.es}}\\\\
{\emph {\small $^3$ Dipartimento di Scienze Fisiche dell'Universit\`a di Napoli ``Federico II'',}}\\
{\emph {\small Complesso Universitario di Monte Sant'Angelo,}}\\
{\emph {\small via Cintia, I-80126 Napoli, Italy}}\\
{\emph {\small E-mail: volkert@na.infn.it}}}

\maketitle

\begin{abstract}
In the case of two qubits, standard entanglement monotones like the linear entropy fail to provide an efficient quantum estimation in the regime of weak entanglement. In this paper, a more efficient entanglement estimation, by means of a novel class of entanglement monotones, is proposed. Following an approach based on the geometric formulation of quantum mechanics, these entanglement monotones are defined by inner products on invariant tensor fields on bipartite qubit orbits of the group $SU(2)\times SU(2)$. 
\end{abstract}




\section{Introduction}

Entanglement is one of the salient features of quantum physics, from the point of view of both foundations \cite{Einstein:1935, bell:87} and emerging quantum technologies \cite{Nielsen:2001, Horodecki:2009}. The empirical evidence of this intriguing phenomenon has been tested by a violation of Bell-type inequalities in nowadays classical experiments \cite{Freedman:1972, Aspect:81}.

In contrast, the task of designing experimental settings able to achieve also \emph{quantitative} estimations of entanglement is a current challenge, even with the most modern quantum experimental devices available today. The fundamental reason may be related to the fact that entanglement monotones are in general non-linear functions on the convex body of quantum states \cite{Vidal:2000}, rather than ordinary expectation values of `quantum observables' associated with Hermitian operators in a Hilbert space.
A possible approach to this problem could be provided, on the one hand, by a tomographic reconstruction of the quantum state involving a series of measurements associated with a complete set of observables (see, e.g., \cite{Ibort:2009} for a review on quantum tomography). On the other hand, while a quantum state is supposed to contain the complete information about a physical system, a sufficient amount of information about entanglement may be extracted by means of an \emph{incomplete} set of observables. This has been illustrated for the concurrence \cite{Wootters:1997id, Coffman:1999jd}, which may be experimentally recovered from four --- rather than fifteen --- parameters associated with the Bloch representation of a bipartite qubit density matrix \cite{Horodecki:2003}. Moreover, an optimal experimental setting for quantifying pure state entanglement --- based on the purity --- which requires the reconstruction of only three parameters associated with local quantum observables, has been proposed \cite{Sancho:2000}.

A general framework for addressing the problem of optimization of experiments is given by suitable \emph{quantum state estimation methods} \cite{Derka:1998, Paris:2004, Paris:2009} originating in the seminal work of Helstrom \cite{Helstrom:1969}, which dates back to the 1960s.

In contrast, the specialization of quantum estimation methods to entanglement is a relatively young field of research (see \cite{Genoni:2008} and references therein). Recent works have focused on the optimization of quantum information measures, like the Kullback mutual information \cite{Acin:2000}, the fidelity \cite{Bagan:2005} and the quantum Fisher information \cite{Genoni:2008}. The latter case has led to concrete experimental realizations \cite{Brida:2010, Brida:2011}.

In particular, the approach based on Fisher information identifies the optimal set of quantum observables by searching for the optimal estimator of parameters associated with a family of quantum states. This identification is possible whenever the quantum Fisher information in the \emph{quantum Cram\`er-Rao inequality} reduces to its classical counterpart \cite{Paris:2009}. Interestingly, in the simplest non-trivial case of a 1-parameter family of bipartite pure qubit states in the Schmidt decomposition, several standard entanglement measure fail to provide an efficient estimation --- in particular --- for \emph{weakly} entangled states \cite{Genoni:2008}.

In this paper, we propose a more efficient estimation in the regime of weak entanglement by taking into account a family of alternative entanglement monotones. Our approach is based on the geometric formulation of quantum mechanics \cite{Strocchi:1956, Cantoni:1975, Cantoni:1977a,Cantoni:1977b,Cantoni:1980,Cantoni:1985,Cirelli:1983,Cirelli:1984, Abbati:1984, Ashtekar:1997ud, Brody:2001, Clemente:2008, Ercolessi:2010}. Indeed, the geometric point of view seems to be quite natural when taking into account the geometric origin of the quantum Fisher information, as provided by the Bures metric and the Fubini-Study metric for mixed and pure states respectively \cite{Uhlmann:1992, Bengtsson:2006, Gibilisco:2007, Facchi:2010}. As a matter of fact, recent developments of the geometric approach concerning finite-level \emph{bipartite} quantum systems \cite{Aniello:09,Volkert:2010iop,Volkert:2010,Aniello:10, Aniello:2011} have produced a tensorial characterization of entanglement by means of the quantum Fisher information, in terms of the pullback of the Fubini-Study metric on submanifolds on orbits induced by the local unitary symmetry group. In particular, it has provided a new class of entanglement monotones candidates related to generalized Poincar\'e invariants \cite{Aniello:10} --- a notion having its origin in geometric methods used in classical mechanics \cite{Arnold:76}. Complementary to these developments, the estimation of pure state entanglement of two qubits \cite{Genoni:2008} can be seen to be provided by the pullback of the Fubini-Study metric on a submanifold that is \emph{not} invariant under local unitary transformations. More specifically, it is a curve intersecting each local unitary orbit at exactly one point. Hence, a geometric approach seems to be suitable to gain a more complete understanding about entanglement monotones and their estimation.

The paper is organized as follows. In section \ref{2} we review the issue of entanglement estimation of two entangled qubits, as recently discussed in \cite{Genoni:2008}. Next, in section \ref{3}, we focus on the corresponding estimation of the purity. Alternative entanglement and purity monotones are introduced in section \ref{4}, using the geometric formulation of quantum mechanics.
After discussing, in section \ref{5}, the estimation of the resulting entanglement and purity measures, a few conclusions are drawn in section \ref{6}.

\section{Local entanglement estimation of two qubits\label{2}}

Consider a 1-parameter family of entangled state vectors
\be \ket{\psi_\lambda} =\sqrt{\lambda}\ket{00}+\sqrt{1-\lambda}\ket{11}, \quad \lambda\in [0,1]\label{ent family} \ee
in a Schmidt decomposition associated with a bipartite quantum system $\H \cong\C^2 \otimes \C^2$ of two pure qubits. We may ask: What is the number $M_{\delta}(\epsilon)$ of measurements required for achieving an optimal estimation of a given entanglement measure 
\be \epsilon: [0,1] \rightarrow [0,1], \quad \lambda\mapsto \epsilon(\lambda) , \label{epsilon}\ee
in a $99,9\%$ confidence interval with fixed relative error $\delta$?
The answer to this question has been given recently in \cite{Genoni:2008}, in the framework of quantum estimation theory \cite{Paris:2009}, by the formula
\be M_{\delta}(\epsilon)= \frac{ \tilde{H}^{-1}(\epsilon)}{\epsilon^2 \delta},\label{Measurements} \ee
where 
\be \tilde{H}(\epsilon):= H(\lambda(\epsilon)) (\partial_{\epsilon}\lambda(\epsilon))^2\label{trafo}\ee
is related to a parameter transformation of the \emph{quantum Fisher information} 
\be H(\lambda):= \rho_{\lambda}(L_{\lambda}^2)\ee
with $L_{\lambda}\in$ End$(\H)$ an operator solving the equation \be \partial_{\lambda}\rho_{\lambda}=[ L_{\lambda},\rho_{\lambda}]_+\ee for a given 1-parameter family of quantum states $\rho_{\lambda}\in D(\H)$. For  the special case of pure states, defined by the normalized rank-1 projectors 
\be \rho_{\lambda}\equiv \ket{\psi_\lambda}\bra{\psi_\lambda},\ee
it is solved by the operator 
\be L_{\lambda}\equiv  2\partial_{\lambda}\ket{\psi_\lambda}\bra{\psi_\lambda}\ee
yielding the quantum Fisher information
\be  H(\lambda)=4\big(\braket{\partial_{\lambda}\p_\lambda}{\partial_{\lambda}\p_\lambda}+\braket{\p_\lambda}{\partial_{\lambda}\p_\lambda}^2\big).\label{H for pure}\ee
For the entangled state vectors (\ref{ent family}) one finds 
\be H(\lambda)=\frac{1}{\lambda(1-\lambda)}.\ee 
Thus the entanglement estimation becomes inefficient for weakly entangled states due to 
\be \lim_{\epsilon\rightarrow 0} M_{\delta}(\epsilon)\rightarrow \infty,\ee
for the (normalized) linear entropy
\be \epsilon(\lambda):=2(1-\Tr((\rho^{A}_{\lambda})^2)=4\lambda(1-\lambda)\ee
and the negativity entanglement measures $\epsilon_N(\lambda):=\sqrt{\epsilon(\lambda)}$ \cite{Genoni:2008}. One may observe in this regard that the linear entropy is directly related to the quantum Fisher information by 
\be \epsilon(\lambda)=\big(\braket{\partial_{\lambda}\p_\lambda}{\partial_{\lambda}\p_\lambda}+\braket{\p_\lambda}{\partial_{\lambda}\p_\lambda}^2\big)^{-1}.\ee 
Actually, the quantum Fisher Information (\ref{H for pure}) for pure states may be linked to the the pullback of the Fubini-Study metric
\be \braket{d\p}{d\p}-\braket{d\p}{\p}\braket{\p}{d\p}\ee
on the 1-parameter family of pure quantum states (see also \cite{Facchi:2010}) according to
\begin{eqnarray}
&\braket{d\p_\lambda}{d\p_\lambda}-\braket{d\p_\lambda}{\p_\lambda}\braket{\p_\lambda}{d\p_\lambda}\label{-}\\& =
\braket{d\p_\lambda}{d\p_\lambda}+\braket{\p_\lambda}{d\p_\lambda}^2\label{+}\\& =\braket{\partial_{\lambda}\p_\lambda}{\partial_{\lambda}\p_\lambda}d\lambda d\lambda+\braket{\p_\lambda}{\partial_{\lambda}\p_\lambda}^2 d\lambda d\lambda,\end{eqnarray}
where we used as a consequence of the normalization $d\braket{\p_\lambda}{\p_\lambda}=0$, i.e.
\be \braket{d\p_\lambda}{\p_\lambda}=-\braket{\p_\lambda}{d\p_\lambda}.\ee
in the transition from (\ref{-}) to (\ref{+}).
In section \ref{4} we will see how the pull-back on the orbits of the symmetry group of entanglement gives rise to a complementary class of monotonic functions.

\section{Local purity estimation of two qubits\label{3}}

The purity
\be \Tr((\rho^{A}_{\lambda})^2)\label{pm} \ee
is complementary to the linear entropy $1-\Tr((\rho^{A}_{\lambda})^2)$. Therefore it becomes natural to ask whether it implies an efficient estimation for weakly entangled states as follows.

The purity measure (\ref{pm}) for the reduced density state \be \rho^{A}_{\lambda}=\left(
\begin{array}{ll}
 \lambda  & 0 \\
 0 & 1-\lambda 
\end{array}
\right)\ee associated with the family of entangled state vectors (\ref{ent family}) reads
\be \epsilon(\lambda):=\Tr((\rho^{A}_{\lambda})^2)= 1-2\lambda+2\lambda^2= \lambda^2+(1-\lambda)^2.\label{purity}\ee
To compute the induced parameter transformation on the quantum Fisher information (\ref{trafo}), we need to identify the inverse function solutions of the purity
\be \lambda(\epsilon):= \frac{1}{2} \left(1\pm\sqrt{2 \epsilon -1}\right).\ee
Both solutions yield according to (\ref{trafo}) the parameter-transformed quantum Fisher information
\be \tilde{H}(\epsilon)=-\frac{1}{4 \epsilon ^2-6 \epsilon +2}.\ee
However, this implies a \emph{negative} number of measurements
\be M_{\delta\equiv 1}(\epsilon)=-4+\frac{6}{\epsilon }-\frac{2}{\epsilon ^2}\label{e of p}\ee
within the estimation (\ref{Measurements}), appearing beyond a physical interpretation. Hence, the purity  does not solve the problem of estimating weakly entangled qubits in an efficient way.
An alternative approach may be provided as follows.

\section{Tensorial characterization of entanglement\label{4}}

According to the geometric formulation of quantum mechanics \cite{Strocchi:1956, Cantoni:1975, Cantoni:1977a,Cantoni:1977b,Cantoni:1980,Cantoni:1985,Cirelli:1983,Cirelli:1984, Abbati:1984, Ashtekar:1997ud, Brody:2001, Clemente:2008, Ercolessi:2010}, one may identify a degenerate covariant tensor field
\be \kappa_{\H_0}:= \frac{d\bar{z}^{j}\otimes dz^{j}}{\sum_j |z^{j}|^2}-\frac{z^{j}d\bar{z}^{j}\otimes \bar{z}^{k}dz^{k}}{(\sum_j |z^{j}|^2)^2}\label{ProjectiveHT}\ee
on a finite-dimensional punctured Hilbert space $\H_0\cong \C^{n+1}-\{0\}$ which is the pull-back of the Fubini study metric induced by the projection on the associated projective space $\C P^{n}$ (if not differently stated, we shall from now on use the Einstein convention by summing over same indices).
The structure (\ref{ProjectiveHT}) decomposes into a real symmetric and an imaginary anti-symmetric part
\be \kappa_{\H_0}:=\eta_{\H_0}+i\omega_{\H_0},\label{deco}\ee relating to a Riemannian and a symplectic structures
on the associated complex projective space respectively.

\subsection{The Pullback on the symmetry group of entanglement}

The pull-back of the Fubini-Study metric on the orbits of the local unitary Lie group
$SU(n)\times SU(n)$
associated with reducible representations on 
 $\H\cong \C^n\otimes \C^n$ has been recently discussed in the decomposition (\ref{deco}) for the characterization of entanglement \cite{Aniello:09,Volkert:2010iop,Volkert:2010,Aniello:10, Aniello:2011}.
We will apply this approach for $n=2$ by using the family of state vectors  (\ref{ent family}) as fiducial vectors for two qubits yielding a $\lambda$-parametrized family of degenerate pullback tensor fields $$\kappa_{SU(2)\times SU(2)}(\lambda):= \eta_{SU(2)\times SU(2)}(\lambda)+i \omega_{SU(2)\times SU(2)}(\lambda)\label{T3}$$
on the Lie group $SU(2)\times SU(2)$. The pullback depends on the choice of a representation of the Lie algebra generators $\{iX_j\}_{j\in \{1,2,..,6\}}$ of $SU(2)\times SU(2)$ on the composite Hilbert space $\H\cong \C^n\otimes \C^n$. By using the product representation \be\{R(X_j)\}_{j\in \{1,2,..,6\}}:=\{\sigma_k\otimes \sigma_0,\sigma_0\otimes \sigma_k\}_{k\in \{1,2,3\}}\ee based on the standard Pauli matrices $\{\sigma_{k}\}_{k\in \{1,2,3\}}$ tensored by the $2\times 2$ identity matrix $\sigma_0$, we shall encounter a decomposition
\be \kappa_{SU(2)\times SU(2)}(\lambda) = \kappa_{jk}(\lambda) \theta^j\otimes \theta^k\label{T1}\ee
into fiducial quantum state dependent (and therefore $\lambda$-dependent) tensor coefficients 
\be \kappa_{jk}(\lambda)= \rho_{\lambda}(R(X_j)R(X_k))-\rho_{\lambda}(R(X_j))\rho_{\lambda}(R(X_k))\label{T2}\ee and a basis of invariant 1-forms $\{\theta^j\otimes \theta^k\}_{j, k\in \{1,2,3\}}$
on $SU(2)\times SU(2)$. The additional decomposition (\ref{T3}) into a real symmetric and imaginary anti-symmetric part yields following geometric interpretation: The symmetric part $\eta_{SU(2)\times SU(2)}(\lambda)$ with the tensor coefficients 
\be \rho_{\lambda}([R(X_j),R(X_k)]_+)-\rho_{\lambda}(R(X_j))\rho_{\lambda}(R(X_k)),\ee encounters as pullback  induced by the projection \be SU(2)\times SU(2) \rightarrow SU(2)\times SU(2)/\mathcal{G}_{0,\lambda}\ee 
on the orbits associated with the isotropy group 
\be  \mathcal{G}_{0,\lambda}:=\{g\in SU(2)\times SU(2)| U(g)\rho_{\lambda}U(g)^{\dagger}=\rho_{\lambda}\}\ee the complete information of a family of non-degenerate pullback Riemannian tensor fields on \emph{all} orbits  (These orbits are classified according to their dimensions 3, 4 and 5 in dependence of maximal entangled ($\lambda=1/2$), separable ($\lambda\in \{0,1\}$) and intermediate  ($\lambda\in (0,1)$) entangled fiducial states $\rho_{\lambda}$ respectively \cite{Kus:2001, Kus:2002}).

The anti-symmetric part
\be\omega_{SU(2)\times SU(2)}(\lambda)=\rho_{\lambda}([R(X_j),R(X_k)]_-)\theta^j\wedge \theta^k\ee  splits --- in  contrast to the symmetric part --- into two families of  tensor fields 
\be\omega_{SU(2)\times SU(2)}(\lambda)= \omega^A_{SU(2)}(\lambda)\oplus \omega^B_{SU(2)}(\lambda).\ee
Each family is defined on a corresponding $SU(2)$-subgroup of $SU(2)\times SU(2)$  by the reduced density state dependent anti-symmetric structures 
\begin{eqnarray}
\omega^A_{SU(2)}(\lambda)  =&\,\rho_{\lambda}([\sigma_j,\sigma_k]_-\otimes \sigma_0)\theta^j\wedge \theta^k\\ =&\rho^A_{\lambda}([\sigma_j,\sigma_k]_-)\theta^j\wedge \theta^k, \end{eqnarray}
and $\omega^B_{SU(2)}(\lambda)=\rho^B_{\lambda}([\sigma_j,\sigma_k]_-)\theta^j\wedge \theta^k$ respectively. 
They become maximal degenerate if and only if the fiducial state is maximal entangled $(\lambda=1/2)$ \cite{Volkert:2010, Aniello:10}. 

\subsection{Monotones from inner products on tensor fields}

All geometric non-degenerate structures described in the previous subsection --- i.e.\,Riemannian tensor fields on orbits of entangled pure states and symplectic tensor fields on submanifolds of reduced density states ---  are completely specified by the coefficients (\ref{T2}) of the $\lambda$-parametrized families of degenerate Hermitian tensor fields
\be \kappa_{SU(2)\times SU(2)}(\lambda) = \kappa_{jk}(\lambda) \theta^j\otimes \theta^k\ee
on $SU(2)\times SU(2)$. These tensor fields are by construction invariant under the symmetry group of entanglement $SU(2)\times SU(2)$ and may therefore naturally mapped to an entanglement monotonic function of $\lambda\in [0,1]$  by considering an Hermitian inner product \begin{eqnarray}\braket{\kappa_{SU(2)\times SU(2)}(\lambda) }{\kappa_{SU(2)\times SU(2)}(\lambda) } := \bar{\kappa}_{jk}(\lambda) \kappa_{rl} (\lambda)\braket{\theta^j\otimes \theta^k}{\theta^r\otimes \theta^l} \end{eqnarray}
on the pullback tensor field $\kappa_{SU(2)\times SU(2)}(\lambda)$ as recently proposed in \cite{Aniello:10}. To keep the formulas as readable as possible, we shall omit in the following the dependency on the parameter $\lambda$ associated the fiducial state $\rho_{\lambda}$ defined in (\ref{ent family}). With \be \braket{\theta^j\otimes \theta^k}{\theta^r\otimes \theta^l}=\braket{\theta^j}{\theta^r}\braket{\theta^k}{ \theta^l}=\delta^{jr}\delta^{kl}\ee one finds then 
\be \braket{\kappa_{SU(2)\times SU(2)}}{\kappa_{SU(2)\times SU(2)}}=
 \bar{\kappa}_{jk} \kappa_{jk}.\label{Trace T2 2}\ee

\subsubsection{Inner products on ordinary tensor products}

\begin{figure}[htp]
\centerline{\includegraphics[scale=0.80]{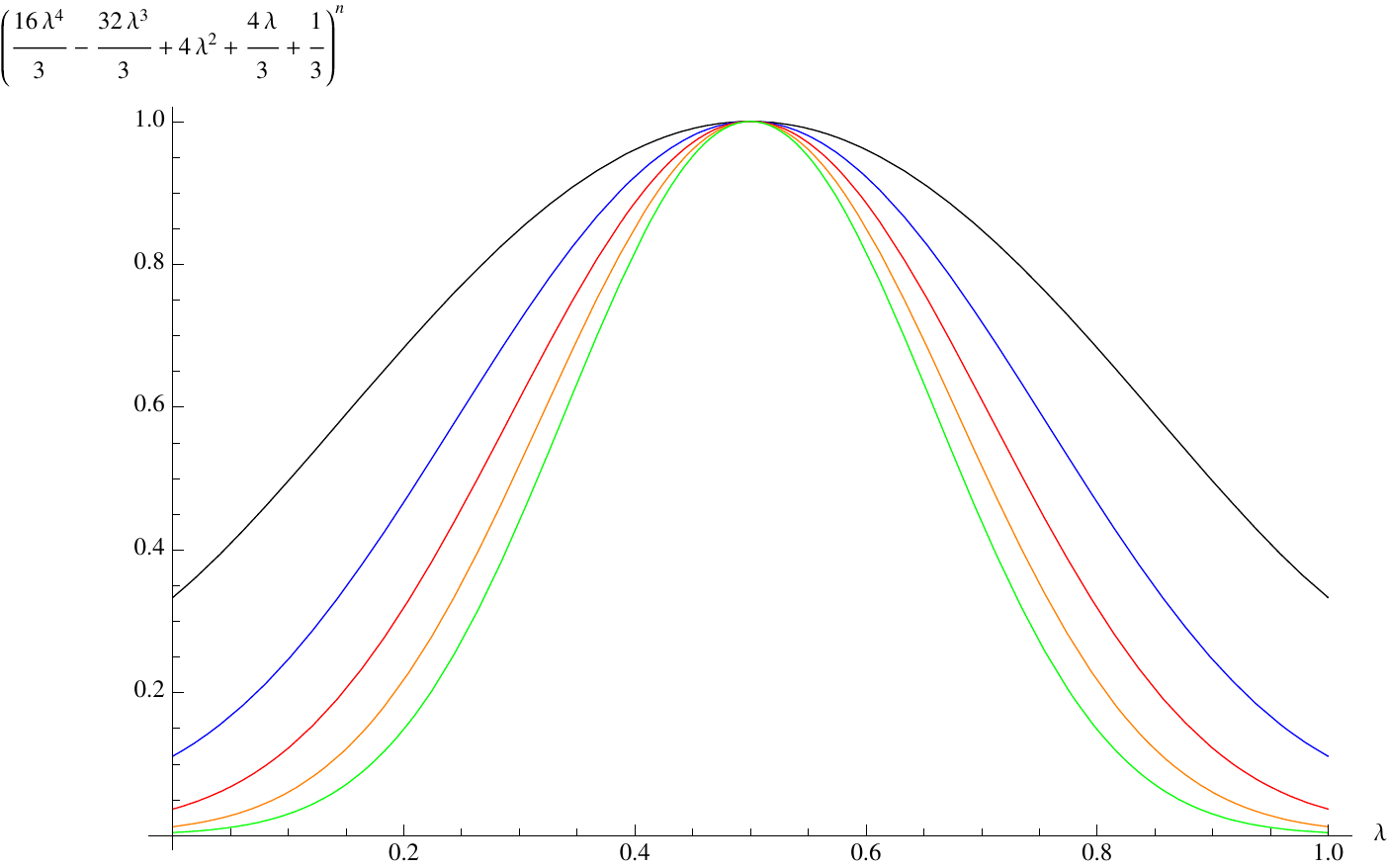}}
\centerline{\includegraphics[scale=1.2]{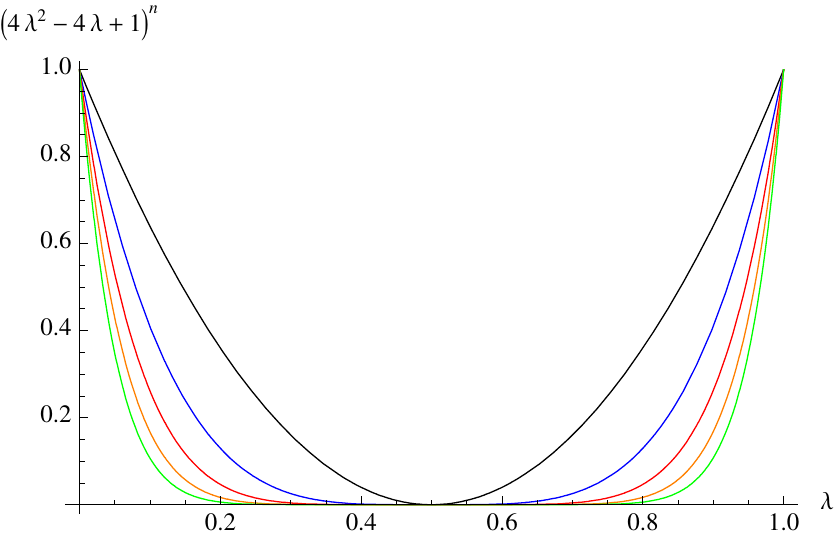}}
\vspace*{8pt}
\caption{A class of entanglement and purity monotones constructed from $SU(2)\times SU(2)$ invariant tensor fields $\eta^{\otimes n}_{SU(2)\times SU(2)}$ and $\omega^{\otimes n}_{SU(2)\times SU(2)}$. The monotones arise here by considering an inner product yielding the invariant functions (\ref{monotone 1}) and (\ref{monotone 2}) respectively. The black curve corresponds in each case to $n=1$ and the following colored curves corresponds to higher order tensor field inner products with $n\in \{2,3,4,5\}$. The entanglement monotones admit in contrast to the purity monotones a normalization in dependence of the tensor order $n$. \label{fig G and O}}
\end{figure}

In the following we may consider a class of  $SU(2)\times SU(2)$-invariant functions arising from higher order tensor fields 
\be \kappa_{SU(2)\times SU(2)}^{\otimes n}:= \bigotimes_{k=1}^n \kappa_{SU(2)\times SU(2)}\ee
by virtue of their corresponding inner product 
\begin{eqnarray} &\braket{\kappa_{SU(2)\times SU(2)}^{\otimes n}}{\kappa_{SU(2)\times SU(2)}^{\otimes n}}= &\braket{\kappa_{SU(2)\times SU(2)}}{\kappa_{SU(2)\times SU(2)}}^n.\label{power on traces}\end{eqnarray}
This can be be proved as follows. The inner product $\braket{T}{T}$ on a general covariant tensor \be  T := T_{j_1 j_2...j_m}\theta^{j_1}\otimes \theta^{j_2}...\otimes \theta^{j_m}\label{order m tensor}\ee
of order $m$ reads
$$  \bar T_{j_1..j_m} T_{k_1..k_m}\braket{\theta^{j_1}\otimes..\otimes \theta^{j_m}}{\theta^{k_1}\otimes ..\otimes \theta^{k_m}}$$
\be  =\bar T_{j_1..j_m} T_{k_1..k_m}\delta_{j_1 k_1}...\delta_{j_m k_m}=\bar T_{j_1..j_m}T_{j_1..j_m}.\label{ip m}\ee
Now, we consider a tensor of even order $m=2n$ constructed from the n-th tensor product of order two tensors
$$  T  \equiv (T_{j_1 j_2}\theta^{j_1}\otimes \theta^{j_2})^{\otimes n}$$ $$= T_{j_1 j_2}T_{j_3 j_4}..T_{j_{m-1} j_m}\theta^{j_1}\otimes \theta^{j_2}\otimes \theta^{j_3}\otimes \theta^{j_4}...\theta^{j_{m-1}}\otimes \theta^{j_m}$$
\be = \prod_{r=1}^{n}  T_{j_{2r-1} j_{2r}} \bigotimes_{r=1}^{n}  \theta^{j_{2r-1}}\otimes \theta^{j_{2r}}.\ee
The tensor coefficients in (\ref{order m tensor}) factorize in this special case (in each term of the sum over same indices) into tensor coefficients of order two
\be  T_{j_1 j_2...j_m}= \prod_{r=1}^{n}  T_{j_{2r-1} j_{2r}}.\ee
Hence, with the inner product (\ref{ip m}) one concludes
$$ \braket{T}{T}=\bar T_{j_1..j_m}T_{j_1..j_m} =  \prod_{r=1}^{n}  \bar{T}_{j_{2r-1} j_{2r}}\prod_{r=1}^{n}  T_{j_{2r-1} j_{2r}}$$\be =\prod_{r=1}^{n}  \bar{T}_{j_{2r-1} j_{2r}}T_{j_{2r-1} j_{2r}} =\prod_{r=1}^{n}  \braket{T}{T} = \braket{T}{T}^n. \ee
This proofs the inner product relation (\ref{power on traces}). 
As a consequence, we may apply the inner product on the tensor product of the symmetric part and the tensor product of the anti-symmetric part separately and find 
\be\braket{\eta_{SU(2)\times SU(2)}^{\otimes n}}{\eta_{SU(2)\times SU(2)}^{\otimes n}}=\braket{\eta_{SU(2)\times SU(2)}}{\eta_{SU(2)\times SU(2)}}^n,\label{sym power on traces}\ee
\be \braket{\omega_{SU(2)\times SU(2)}^{\otimes n}}{\omega_{SU(2)\times SU(2)}^{\otimes n}}=\braket{\omega_{SU(2)\times SU(2)}}{\omega_{SU(2)\times SU(2)}}^n.\label{a sym power on traces}\ee
The inner products (\ref{sym power on traces}) and (\ref{a sym power on traces}) applied here to the higher order tensor products of pullback tensor fields associated with the family of Schmidt decomposed fiducial states (\ref{ent family}) establish the main result of the present section: We  find \emph{entanglement monotones} for the inner product on the tensor products of the symmetric part and \emph{purity monotones} for the inner product on the tensor products of the anti-symmetric part as illustrated within an appropriate normalization in figure \ref{fig G and O} for the first five tensor power orders. 
As normalization we use in this regard the factors
\be \frac{1}{(2\mbox{Dim}(SU(2)\times SU(2))^n}  = \frac{1}{12^n} ,\ee
for the inner products (\ref{sym power on traces}) and the factors
\be \frac{1}{\mbox{Dim}(S^2\times S^2)^n}=\frac{1}{4^n}.\ee
for the inner products (\ref{a sym power on traces}). 
The resulting functions are higher order polynomials on the Schmidt coefficient $\lambda$
\be \frac{1}{12^n} \braket{\eta_{SU(2)\times SU(2)}^{\rho_{\lambda}}}{\eta_{SU(2)\times SU(2)}^{\rho_{\lambda}}}^n= \left(\frac{16 \lambda ^4}{3}-\frac{32 \lambda ^3}{3}+4 \lambda
   ^2+\frac{4 \lambda }{3}+\frac{1}{3}\right)^n\label{monotone 1}\ee
\be \frac{1}{4^n}\braket{\omega_{SU(2)\times SU(2)}^{\rho_{\lambda}}}{\omega_{SU(2)\times SU(2)}^{\rho_{\lambda}}}^n= \left(4 \lambda ^2-4 \lambda +1\right)^n.\label{monotone 2}\ee
establishing a \emph{quantitative} justification for the identification between \emph{maximal entangled states} and \emph{Lagrangian entangled} states $(\lambda=1/2)$ on the value `1' for the entanglement monotones and the value `0' for the purity monotones respectively. 
The purity monotone is normalized for separable states ($\lambda\in\{0,1\}$) to `$1$'. In contrast, we find in the entanglement monotone a normalization
\be \frac{1}{12^n} \braket{\eta_{SU(2)\times SU(2)}^{\rho_{\lambda}}}{\eta_{SU(2)\times SU(2)}^{\rho_{\lambda}}}^n|_{\lambda\in\{0,1\}}=\frac{1}{3^n},\ee
being \emph{dependent on the tensor field order} $n$. We may therefore recover for $n \rightarrow \infty$  the `standard normalization' for separable states. This indicates that the inner products on $n$-order tensor products of the symmetric tensor field provide an \emph{approximation} to a `bona fide' entanglement measure with increasing $n$. This approximation will provide an advantage in contrast to standard entanglement measures when testing their quantum estimation efficiency as shown later on in section \ref{5}.  

\subsubsection{Inner products on symmetric tensor products} 

So far we considered entanglement-invariants arising from inner products on ordinary tensor products on pullback tensor fields associated with $SU(2)\times SU(2)$-orbits.  By considering now \emph{symmetrized} tensor products, we may find a distinguished class of invariants. Such an approach may be seen complementarily related to the notion of Poincar\'e invariants which are constructed from higher order anti-symmetric structures $\omega^{\wedge n}:=\omega \wedge \omega...\wedge \omega $ \cite{Arnold:76}. Starting from the symmetric part of the pullback tensor field
$\kappa = \eta + i \omega$
we define here
\be \eta^{\vee  n}:=\eta \vee  \eta...\vee\eta  \ee
and compute  
$$ \braket{\eta_{SU(2)\times SU(2)}^{\vee  n}}{\eta_{SU(2)\times SU(2)}^{\vee  n}}=$$
$$\prod_{r=1}^{n}  \eta_{j_{2r-1} j_{2r}}  \eta_{k_{2r-1} k_{2r}}
\braket{ \bigvee_{r=1}^{n}\theta^{j_{2r-1}}\vee\theta^{ j_{2r}}}{\bigvee_{r=1}^{n}\theta^{k_{2r-1}}\vee\theta^{ k_{2r}}} $$
The inner product on the $2n$-th symmetric product of 1-forms $\theta^{k_i}$ involves the \emph{permanent} of a matrix with Kronecker deltas $\delta^{j_{r}k_{r}}$ according to
\be \braket{ \bigvee_{r=1}^{2n}\theta^{j_{r}}}{\bigvee_{r=1}^{2n}\theta^{k_{r}}} = \frac{1}{2n!} \mbox{per}\big(\braket{\theta^{j_{r}}}{\theta^{k_{r}}}\big).\ee
The notion of the \emph{permanent} of a matrix may be familiar from the inner product as used for `bosons' (see e.g.\cite{Grabowski:2011}). Indeed, we shall underline the distinguished physical interpretation in our setting: We are dealing with \emph{distinguishable} particles associated with an ordinary tensor product Hilbert space together with symmetrized `classical' higher order tensor fields as \emph{additional} structure on a Lie group manifold.

\begin{figure}[htp]
\centerline{\includegraphics[scale=1.10]{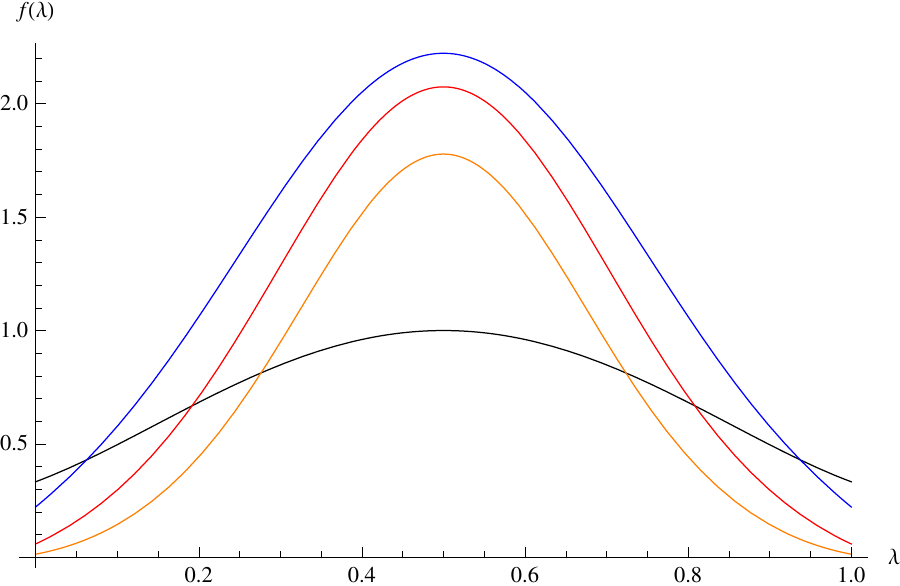}}
\vspace*{8pt}
\caption{A class of entanglement monotones constructed from \emph{symmetric} products of symmetric $SU(2)\times SU(2)$ invariant tensor fields $\eta^{\vee n}_{SU(2)\times SU(2)}$. 
The black curve corresponds in each case to $n=1$ and the following colored curves corresponds to higher order symmetric tensor field inner products with $n\in \{2,3,4\}$. \label{fig symInv}}
\end{figure}

The expression of the inner product 
$$ \braket{\eta_{SU(2)\times SU(2)}^{\vee  n}}{\eta_{SU(2)\times SU(2)}^{\vee  n}}=
\prod_{r=1}^{n}  \eta_{j_{2r-1} j_{2r}}  \eta_{k_{2r-1} k_{2r}}\frac{1}{2n!} \mbox{per}\big(\delta^{j_{r}k_{r}}\big) $$
becomes cumbersome with increasing tensor product orders. To reduce the computational effort we perform a basis transformation in which the tensor coefficient matrix associated with $\eta$ becomes diagonal. In this way we find up to a normalization,  the following resulting invariants for the first three orders with $n=1$:
\be \,\, 16 \lambda ^4-32 \lambda ^3+12 \lambda ^2+4 \lambda +1,\ee
$n=2:$
$$\frac{2}{27} (1536 \lambda ^8-6144 \lambda ^7+8960 \lambda ^6  -5376 \lambda ^5$$
\be +880 \lambda ^4+32 \lambda
   ^3+72 \lambda ^2+40 \lambda +3),\ee
and $n=3:$
$$\frac{8}{135}(20480 \lambda ^{12}-122880 \lambda ^{11}+304128 \lambda
   ^{10}$$ $$-394240 \lambda ^9+277632 \lambda ^8-96768 \lambda ^7+11648 \lambda ^6$$
\be-384\lambda ^5+136 \lambda ^4+112 \lambda ^3+108 \lambda ^2+28 \lambda +1).\ee
The corresponding graphs are illustrated together with $n=4$ in figure \ref{fig symInv}.

\section{Estimation of inner products on tensor fields \label{5}}

In the following we ask: What happens if we estimate the entanglement monotones provided by the inner products 
\bea \epsilon_n :=& \frac{1}{12^n}\braket{\eta_{SU(2)\times SU(2)}^{\otimes n}}{\eta_{SU(2)\times SU(2)}^{\otimes n}}  \\ \mu_n :=& \frac{(-1)^n}{4^n}\braket{\omega_{SU(2)\times SU(2)}^{\otimes n}}{\omega_{SU(2)\times SU(2)}^{\otimes n}} \eea
on the tensor products of the symmetric and anti-symmetric part of the pullback tensor fields
\be \kappa_{SU(2)\times SU(2)}= \eta_{SU(2)\times SU(2)}+i \omega_{SU(2)\times SU(2)} \ee
as constructed in the previous section?

\begin{figure}[htp]
\centerline{\includegraphics[scale=1.00]{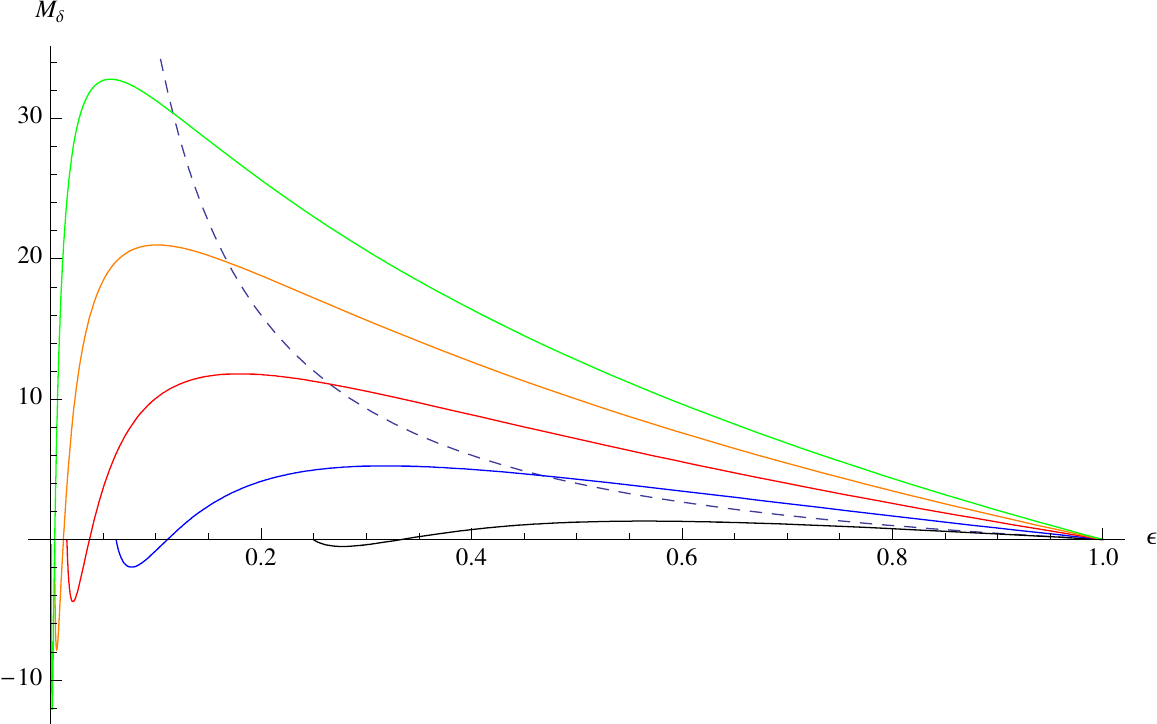}}
\vspace*{8pt}
\centerline{\includegraphics[scale=1.20]{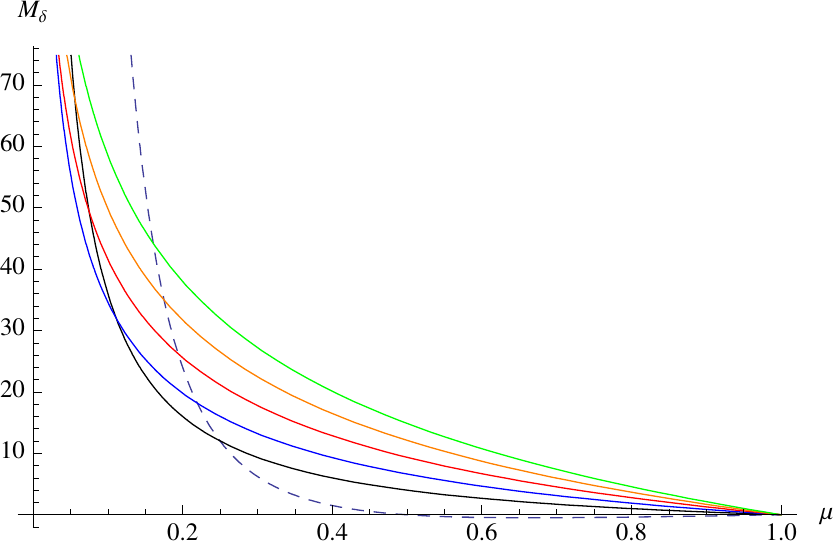}}
\vspace*{8pt}
\caption{estimation of entanglement monotones $\epsilon_n$ and purity monotones $\mu_n$ for the first five orders (each order corresponds to the same color as used in the plot in figure \ref{fig G and O}). It shows the number $M_{\delta}$ of measurements in dependence of the value $\epsilon_n$ (and $\mu_n$ respectively) of the monotones  required for achieving an estimation in a $99,9\%$ confidence interval with fixed relative error $\delta\equiv 1$. The dashed curves in the first plot correspond to the estimation of the linear entropy, and in the second plot to the negative-valued estimation of the purity as done in (\ref{e of p}) when `reflected' on the $\mu$-axis. \label{fig a}}
\end{figure}

For this purpose we apply the procedure of section \ref{2} by first identifying the inverse functions
$\lambda(\epsilon_n)$ and $\lambda(\mu_n)$ of the monotones (\ref{monotone 1}) and (\ref{monotone 2}),
\be \left(\frac{16 \lambda ^4}{3}-\frac{32 \lambda ^3}{3}+4 \lambda
   ^2+\frac{4 \lambda }{3}+\frac{1}{3}\right)^n\ee
\be    \left(4 \lambda ^2-4 \lambda +1\right)^n\ee
 as solutions of the equations
\be \left(\frac{16 \lambda ^4}{3}-\frac{32 \lambda ^3}{3}+4 \lambda
   ^2+\frac{4 \lambda }{3}+\frac{1}{3}\right)^n -\epsilon_n(\lambda)=0\label{sol monotone 1}\ee
\be  \left(4 \lambda ^2-4 \lambda +1\right)^n-\mu_n(\lambda)=0\label{sol monotone 2}\ee
associated with the inner products of the tensor products of the symmetric tensor fields and the tensor product of the antisymmetric tensor fields respectively. As a result we find both real and imaginary valued solutions. To provide a physical interpretation we consider the real-valued solutions and define the parameter transformation  \be \kappa_{\epsilon_n}:= \kappa_{\lambda(\epsilon_n)} (\partial_{\epsilon_n}\lambda(\epsilon_n))^2\ee
of the quantum Fisher information on the 1-parameter family of Schmidt coefficient decomposed quantum states  
according to formula (\ref{trafo}). In this way we find the minumum 
 number $M_{\delta}(\epsilon_n)$ of measurements \be M_{\delta}(\epsilon_n)= \frac{1}{\epsilon^{2}_n \delta^{2}} \frac{1}{ \kappa_{\epsilon_n}} \ee as defined in (\ref{Measurements}) for achieving an estimation with fixed relative error $\delta$. The result is illustrated with $\delta \equiv 1$ for the first five powers in figure \ref{fig a}. 

\subsection{Discussion for monotones from the symmetric part}

Let us begin to analyze the estimation of the entanglement monotones associated with the inner products on the tensor products of the symmetric tensor field in the first plot of figure \ref{fig a}. While the estimation of the linear entropy (dashed curve) diverges for weakly entangled states in the limit $\epsilon_n \rightarrow 0$ (see section \ref{2} and \cite{Genoni:2008}), we find for the tensorial monotones an approximative improvement with a finite number of measurements towards the regime of weak entanglement in dependence of the tensor field order $n$. The inflection point into negative values indicates the boundary of the regime where the approximation looses its validity.  The validity of the approximation into the regime of weekly entanglement may become enlarged by considering inner products on tensor fields of higher order. The green curve in the first plot of figure \ref{fig a} corresponds to the highest order example $n=5$ and clearly illustrates this enlargement when compared to the lower tensor field orders.

\subsection{Discussion for monotones from the anti-symmetric part}

The estimation of the purity monotones associated with inner products on the tensor products of the anti-symmetric tensor fields is illustrated in the second plot of figure \ref{fig a}. All curves clearly show here an efficient estimation for weakly entangled states, i.e.\,for all states close to $\mu_n = 1$.  The dashed curve corresponds here to the estimation of the standard purity (section \ref{3}) when reflected on the $\mu$-axis into positive values. Indeed, only the curves associated with the   inner products on the tensor fields may admit a physical interpretation.

\section{Conclusions and outlook\label{6}}

In the geometric formulation of quantum mechanics one considers the Fubini-Study metric at the first place. Any action of the symmetry group of entanglement on a family of entangled quantum state vectors induces a family of degenerate pull back tensor fields each defined as pull back of the Fubini-Study metric from the Hilbert space to the Lie group $SU(2)\times SU(2)$. Along the decomposition of the Fubini-Study metric into a Riemannian and a symplectic tensor field
one finds a decomposition $$\kappa_{SU(2)\times SU(2)} = \eta_{SU(2)\times SU(2)} +i\omega_{SU(2)\times SU(2)} $$ into degenerate symmetric and anti-symmetric pullback structures. Via an inner product on higher order tensor fields it is possible to identify two  classes of monotonic functions characterizing the entanglement and purity of a bipartite quantum system. These geometrically constructed classes of entanglement and purity monotones provide advantages in the estimation of entangled qubits when compared to standard entanglement and purity monotones. The basic picture emerging here may be subsumed as follows. While the inner product  
$$\braket{\eta_{SU(2)\times SU(2)}^{\otimes n}}{\eta_{SU(2)\times SU(2)}^{\otimes n}}$$ 
yields an \emph{approximative} efficient entanglement estimation for \emph{all} state vectors, one finds for the inner product 
$$ \braket{\omega_{SU(2)\times SU(2)}^{\otimes n}}{\omega_{SU(2)\times SU(2)}^{\otimes n}}$$ an \emph{exact} efficient purity estimation for \emph{weakly} entangled state vectors.

It would be interesting to investigate whether this approach admits also advantages in the entanglement estimation of more general composite quantum systems involving multi-partite systems, mixed quantum states and infinite dimensional Hilbert spaces. As a matter of fact, such a generalization becomes directly testable by the tensor field-valued pairing  
\bea & \{\rho_{\vec{\lambda}}\}_{\vec{\lambda}\in \mathcal{M}}  \times \{U^{-1}(g)dU(g)^{\otimes k}\}_{g \in \mathcal{G}}\\  &\downarrow\\ &  \rho_{\vec{\lambda}}\bigg(U^{-1}(g)dU(g)^{\otimes k}\bigg)  \equiv  \kappa_{\mathcal{G}}(\vec{\lambda}) \eea
as described in \cite{Aniello:10} between general manifolds $\{\rho_{\vec{\lambda}}\}_{\vec{\lambda}\in \mathcal{M}}$ of quantum states and \emph{invariant operator valued tensor fields} $U^{-1}(g)dU(g)^{\otimes k}$ on general Lie groups $\mathcal{G}$ associated with a unitary representation $U:\mathcal{G}\rightarrow \mathcal{U}(\H)$.
A generalization in several directions could therefore be tackled by focusing on the corresponding    inner products on tensor fields  
$$\braket{\kappa_{\mathcal{G}}^{\otimes n}(\vec{\lambda})}{\kappa_{\mathcal{G}}^{\otimes n}(\vec{\lambda})}.$$  
Finally, a deeper understanding of the relation of our results with the quantum Fisher information (see \cite{Gibilisco:2007} and references therein) may be achieved in terms of a suitable geometrization of the $C^*$-algebraic approach to quantum mechanics \cite{Carinena:2007ws, Chruscinski:2008}, including the formalism of star products of quantum tomograms \cite{Ibort:2009}, which may close a circle to the empirical bounds on the precision of quantum measurements in terms of generalized uncertainty relations \cite{Man'ko:2002}.

\section*{Acknowledgments}
This work has been supported by the National Institute of Nuclear Physics (INFN).

\bibliography{mybib}

\end{document}